\documentclass[aps,prl,twocolumn,showpacs,groupedaddress]{revtex4-1}
\usepackage{graphicx}% Include figure files
\usepackage{dcolumn}% Align table columns on decimal point
\usepackage{bm}% bold math
\begin{document}

%\preprint{AIP/123-QED}

\title{Experimental Evidence of Directivity-Enhancing Mechanisms in Nonlinear Lattices}

\author{R. Ganesh}
\email{ramak015@umn.edu}
\author{S. Gonella}%
\email{sgonella@umn.edu}
\affiliation{Department of Civil, Environmental, and Geo- Engineering, University of Minnesota Minneapolis, Minnesota, 55455, USA}%

\date{\today}% It is always \today, today,

\begin{abstract}
In this letter, we experimentally investigate the directional characteristics of propagating, finite-amplitude wave packets in lattice materials, with an emphasis on the functionality enhancement due to the nonlinearly-generated higher harmonics. To this end, we subject a thin, periodically perforated sheet to out-of-plane harmonic excitations, and we design a systematic measurement and data processing routine that leverages the full-wavefield reconstruction capabilities of a laser vibrometer to precisely delineate the effects of nonlinearity. % we design a sophisticated measurement and processing routine that fully leverages the measurement sensitivity and sensing flexibility of the laser vibrometer.% by utilizing novel techniques to realize amplitude-dependent features, %elementary signal processing techniques to analyze the experimentally-reconstructed full wavefield data, 
We demonstrate experimentally that the interplay of dispersion, nonlinearity, and modal complexity which is involved in the generation and propagation of higher harmonics gives rise to secondary wave packets with characteristics that conform to the dispersion relation of the corresponding linear structure. Furthermore, these nonlinearly generated wave features display modal and directional characteristics that are complementary to those exhibited by the fundamental harmonic, thus resulting in an augmentation of the functionality landscape of the lattice. %Moreover, they activate mechanisms typical of high-frequency regimes, even while operating at low frequencies of excitation. 
These results provide proof of concept for the possibility to engineer the nonlinear wave response of mechanical metamaterials through a geometric and topological design of the unit cell.
\end{abstract}

% PACS, the Physics and Astronomy Classification Scheme.
\maketitle

Periodic structures and materials feature an intrinsic ability to impede wave propagation within certain frequency ranges referred to as bandgaps \cite{Sigalas1995, Liu2000}, which allows them to effectively function as vibration filters and waveguides \cite{Matlack2016,CelliLego}. Another related, but significantly less explored effect is their frequency-dependent spatial directivity \cite{Phani2006}, by which propagating wave packets travel with anisotropic characteristics and display patterns whose morphology depends on the frequency of excitation. When periodic structures experience finite-deformations, both bandgaps and directivity become amplitude-dependent \cite{Narisetti2010, Narisetti2011}, thereby endowing the structure with the ability to adapt to changes in operating conditions. On the other hand, nonlinear periodic structures can also be tuned to elicit complementary responses without changes in the operating condition \cite{Wang2014}, a feature that is commonly exploited to design tunable vibration filters \cite{Boechler2011a}. Finally, the availability of nonlinear mechanisms also results in the ability to realize devices such as acoustic diodes and rectifiers \cite{Boechler2011, Devaux2015, Neel2016}, by triggering non-reciprocal effects that are seldom possible in their linear counterparts. \newline
The most well-known signature of finite-amplitude wave propagation is the generation of harmonics \cite{DeLima2003}, a feature which, in conjunction with dispersion, is commonly employed as an inspection and characterization tool in Non-Destructive Evaluation (NDE) techniques \cite{Deng2005, Deng2007, Bermes2007}. While the concept of harmonic generation has also been explored in nonlinear periodic structures \cite{Goncalves2000, Cabaret2012}, its complete implications on the spatial characteristics of wave propagation have only been marginally studied. In this regard, we recently carried out a theoretical and numerical investigation of granular phononic crystals \cite{Ganesh2015} and nonlinear lattices \cite{Ganesh2016}, where we have shown that the onset of nonlinear mechanisms (obtained, for example, by  increasing the amplitude of excitation) can be effectively used to stretch the frequency signature of the wave response and distribute it over multiple modes, thereby activating a mixture of (possibly complementary) modal characteristics and enabling functionalities associated with high-frequency optical modes, even while operating in the low-frequency regime. In this letter, we provide a first experimental proof of concept of these effects using a simple lattice geometry to demonstrate the potential of nonlinearity as a tool to attain engineering materials with desired tunable and switchable functionalities. \newline
\begin{figure}[!htb]
\centering
\includegraphics[scale=0.135]{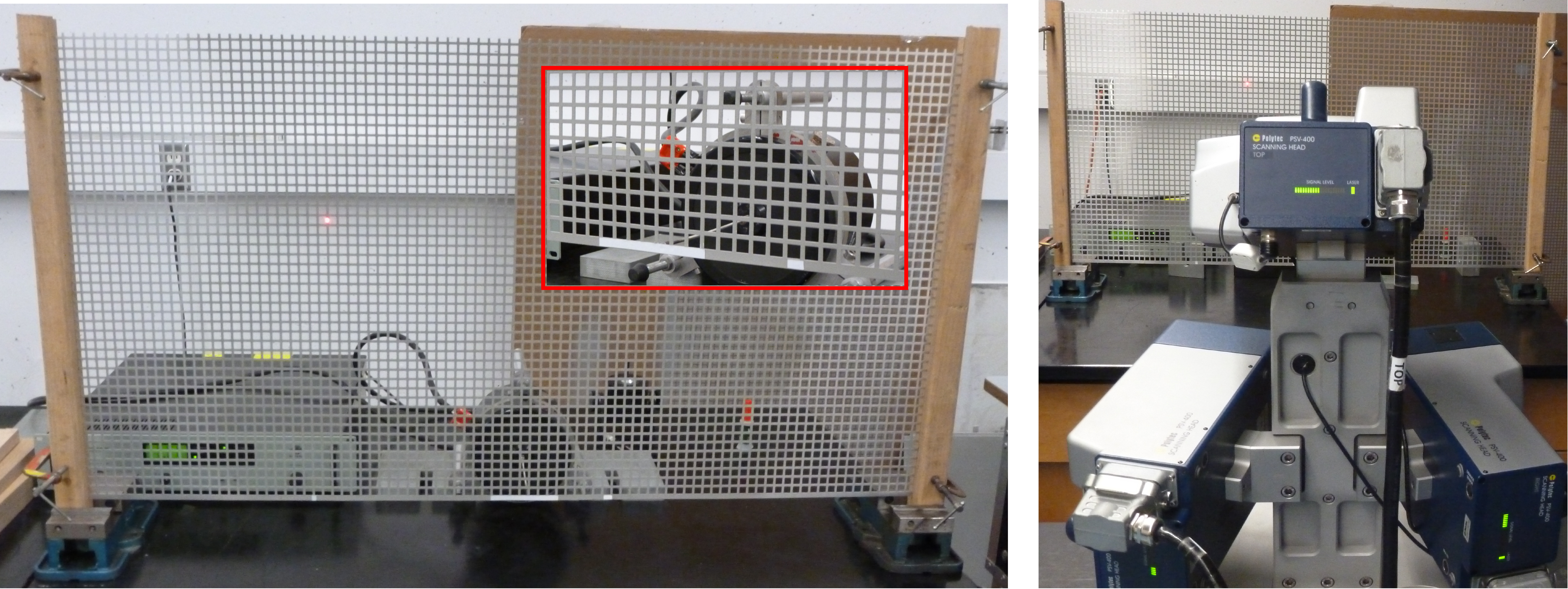}
\caption{Experimental setup (with the source of excitation shown in the inset) and the $3$D Scanning Laser Vibrometer.}
\label{fig1}
\end{figure}
The structure under investigation is a periodically perforated, thin Aluminum sheet (shown in fig.~\ref{fig1}), obtained pre-cut to a size of $4'$($1.2192$ m)$\,\times\,2'$($0.6096$ m) from McNICHOLS (Part No.~$1796003241$). The perforations are $0.375"\,$($9.525$ mm) square holes, spaced $0.5"\,$($12.7$ mm) apart in the horizontal and vertical directions, and the out-of-plane thickness of the sheet is $0.032"\,$($0.8128$ mm). The large size of the structure is motivated by the interest in eliminating the effects of boundary conditions on wave propagation (contamination of the wavefield due to reflections from the edges). The choice to work with Aluminum was supported by the fact that nonlinear effects (attributed to the cubic order terms in the strain energy) have been experimentally observed in thin Aluminum plates experiencing Lamb waves\cite{Deng2007, Matlack2011}. In these experiments, the nonlinear terms in the response are very small when compared to the linear effects, in compliance with the fundamental assumptions of weak nonlinearity that underpin our analysis. \newline%is based on the assumption that the magnitude of wave propagation is small with respect to the length scale of the structure under consideration, albeit large enough to elicit the effects of finite-deformation,
The dispersion relation of the square lattice is determined by considering a linearized model of the structure (the quadratic nonlinearity leads to a very small correction of the dispersion relation, which can be neglected\cite{Ganesh2013}), and following a standard finite element based Bloch analysis procedure using $3$D isoparametric brick elements\cite{Phani2006}.%of the structure is sufficiently captured by that of the linearized model, and is estimated by following a standard procedure in phononics analysis\cite{Phani2006}, and is shown in fig.~\ref{fig2} (the unit cell of the periodic structure is a $3$D square lattice, with the tessellation only in the X-Y plane).
\begin{figure}[!htb]
\includegraphics[scale=0.235]{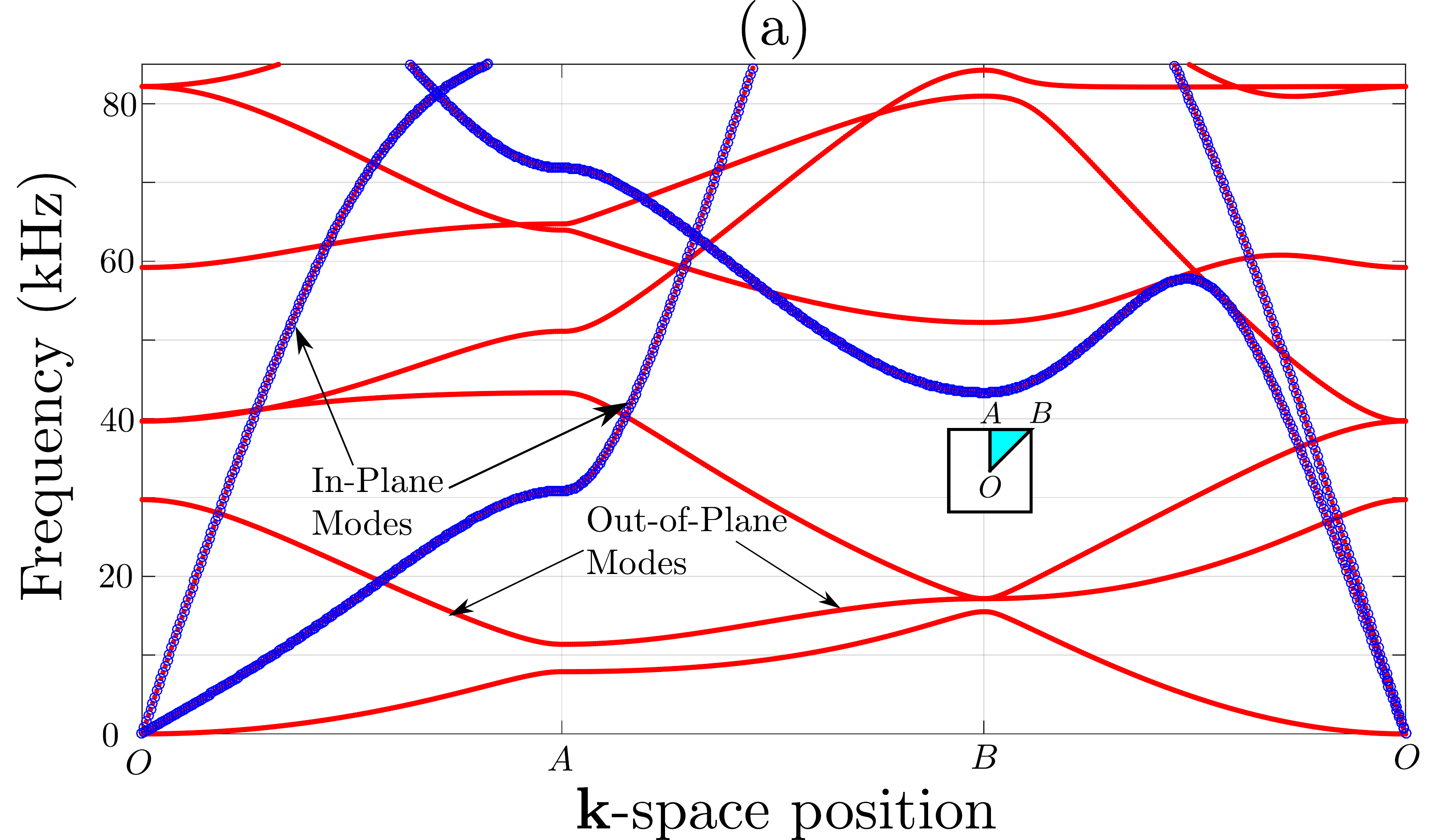} \\\vspace{0.05in}
\includegraphics[scale=0.235]{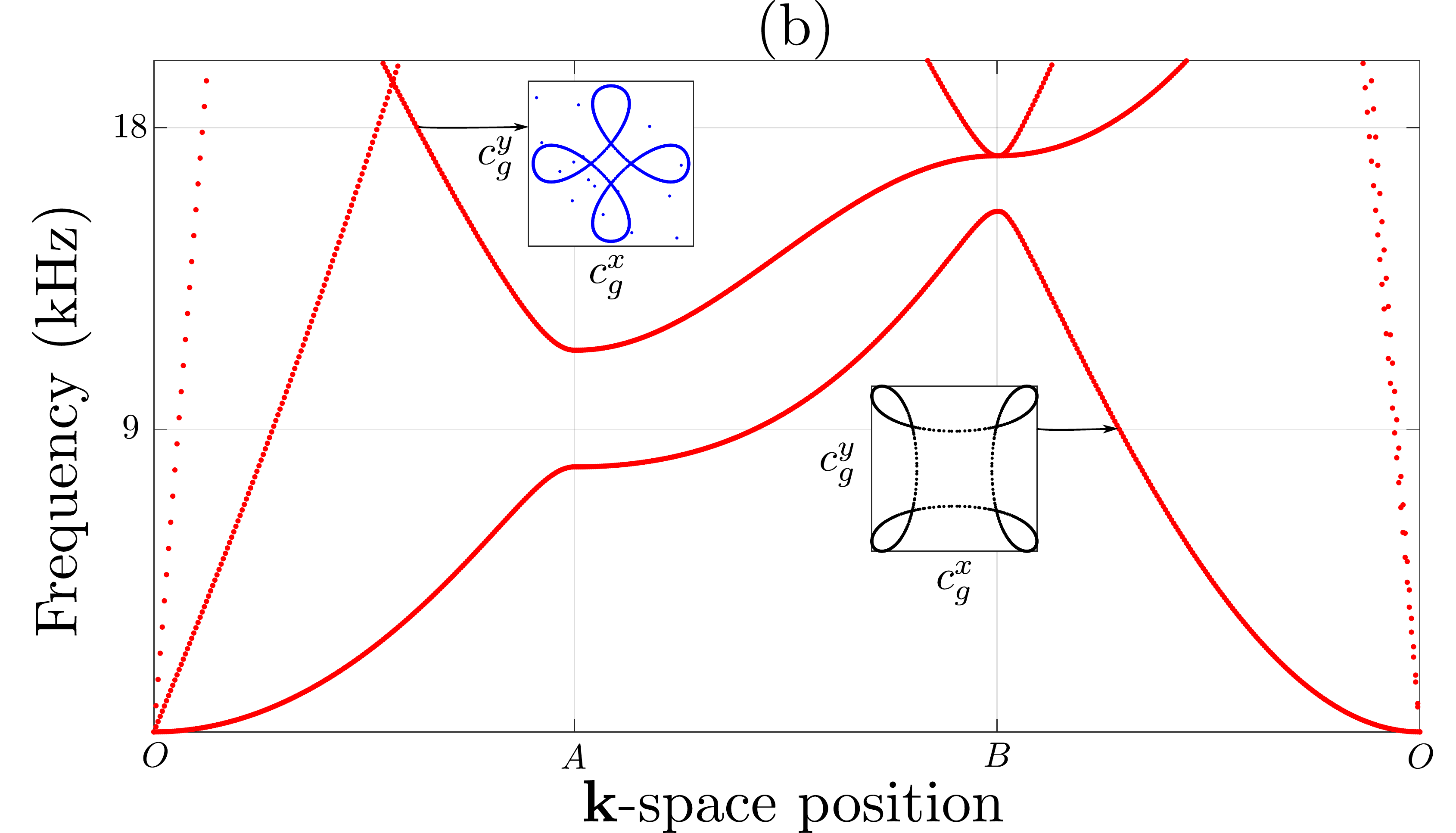}		   
\caption{(a) Linearized dispersion relations obtained using a 2D and 3D model of the unit cell to delineate the in-plane and out-of-plane modes of wave propagation. (b) Magnified details of the first two out-of-plane modes. The group velocity contours at the two selected frequencies are shown in the insets.}
\label{fig2}
\end{figure}
The band structure shown in fig.~\ref{fig2}(a) consists of a mix of in-plane and out-of-plane wave modes. In order to distinguish between the two families of modes, Bloch analysis is repeated using a $2$D plane stress model of the unit cell to determine the band structure of the in-plane modes independently. By superimposing the results of the $2$D and the $3$D analysis, the dispersion curves for the out-of-plane modes can be inferred, as illustrated in fig.~\ref{fig2}(a). Since the thickness of the sheet is much smaller than the other dimensions, the out-of-plane response has a higher likelihood of exhibiting finite-deformation effects, even for relatively small amplitudes of excitation. Therefore, we focus our attention on out-of-plane flexural wave motion. \newline 
We consider two particular frequencies ($\omega = 9\,$kHz and $\omega = 18\,$kHz) lying on the first and second out-of-plane modes respectively, as they appear to be well-suited to investigate modal mixing effects (fig.~\ref{fig2}b). The spatial directivity of the lattice at these frequencies is determined from the group velocity contours, which are depicted in fig.~\ref{fig2}(b) for the selected frequencies. These contours predict the existence of preferential directions of propagation which, in this structure, correspond to the diagonal directions for an excitation at $9$ kHz, and to the horizontal and vertical directions for an excitation at $18$ kHz. The complementary directional patterns of the two frequencies accentuate our ability to investigate the spatial characteristics of finite-amplitude traveling wave packets in this structure. \newline
We excite the lattice by applying a seven-cycle Hann-modulated tone-burst excitation at the center of the bottom edge (in the out-of-plane direction, as shown in the inset in fig.~\ref{fig1}) using an electrodynamic shaker (Bruel \& Kjaer Shaker Type $2809$), and we measure the response using a $3$D Scanning Laser Doppler Vibrometer (SLDV, Polytec PSV-$400$-$3$D). Laser Doppler Vibrometry (LDV) is a non-contact measurement technique which exploits high fidelity lasers and the Doppler effect to measure the velocity of a vibrating object, and has been recently employed to sense, and spatially reconstruct propagating waves in lattice structures \cite{Celli2014, Andreassen2014, Xiao2014, CelliLego}. % Since the velocity of the vibrating object is determined along the line of incidence of the laser, a $3$D LDV utilizes concurrent measurement by three lasers at a single point on the object of interest to compute the complete vibration characteristics of the chosen point. Additionally, the $3$D LDV system supplied by Polytec Inc. (PSV $400$-$3$D) features the capability to sequentially determine the vibrations at multiple points of interest on the object, thereby rendering it the ability to scan and measure the response of the entire object. 
Since the accuracy of the measurements depend on the quality of the laser beam reflected back from the target to the scanning head, the Aluminum sheet is coated with a layer of reflective paint, and additional patches of reflective tape are also placed at critical measurement locations. Furthermore, the accuracy of the $3$D SLDV measurements also depend on the ability to direct all the three lasers concurrently at the same point on the object. This is achieved by an in-built video triangulation process, where the coordinates of the scan points supplied to the system prior to the measurement are used to direct all the laser beams such that they converge on to each scan point. Since the coalescence of the laser beams may fail at a few locations on the measurement surface, a redundant number of scan point definitions are supplied to the laser vibrometer and, after acquisition, only the scan points with valid data are retained for the analysis. Finally, each measurement is repeated multiple times, and a time-average is performed to minimize the noise-to-signal ratio. \newline 
\begin{figure*}[!htb]
\includegraphics[scale=0.4]{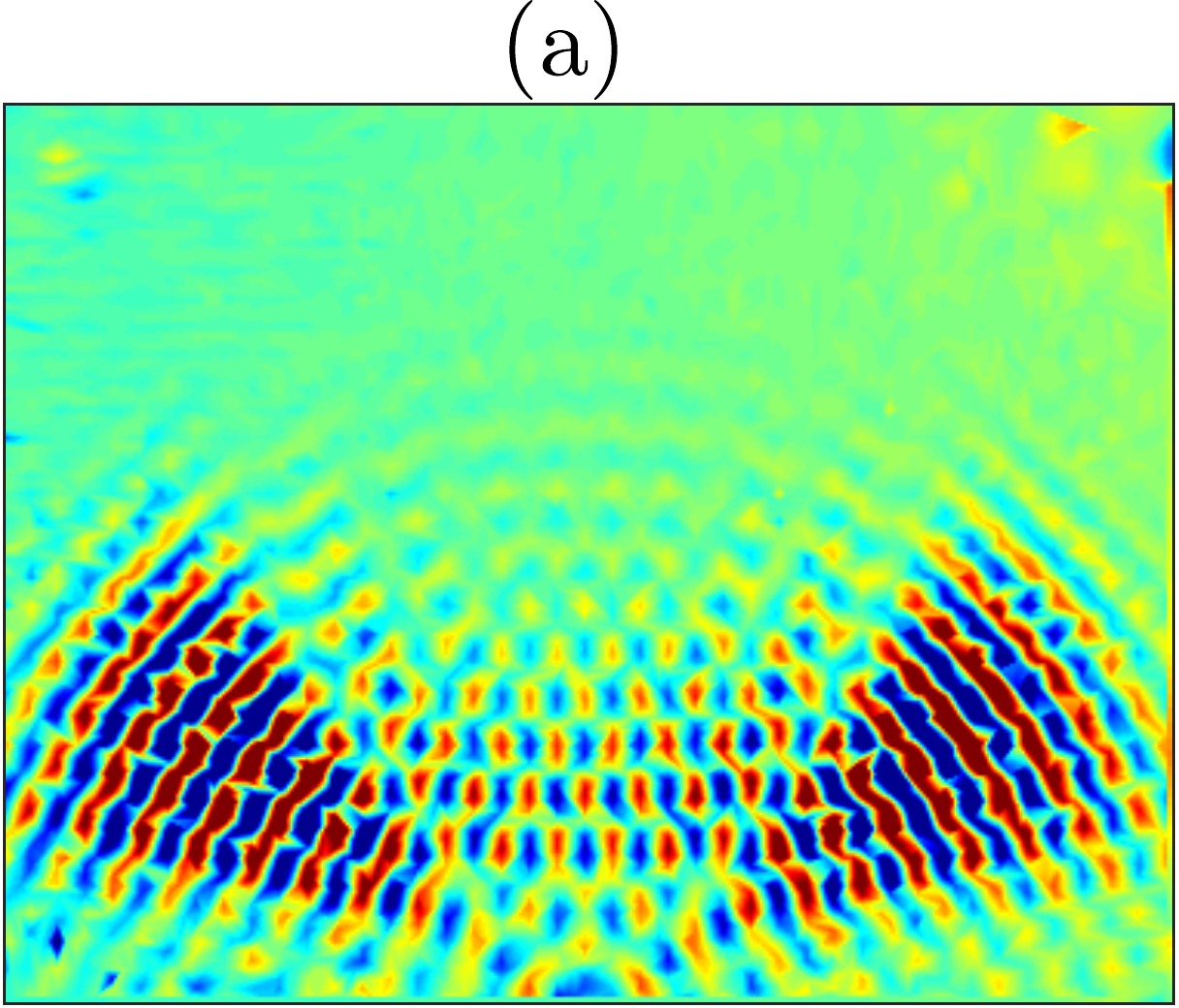} \qquad
\includegraphics[scale=0.4]{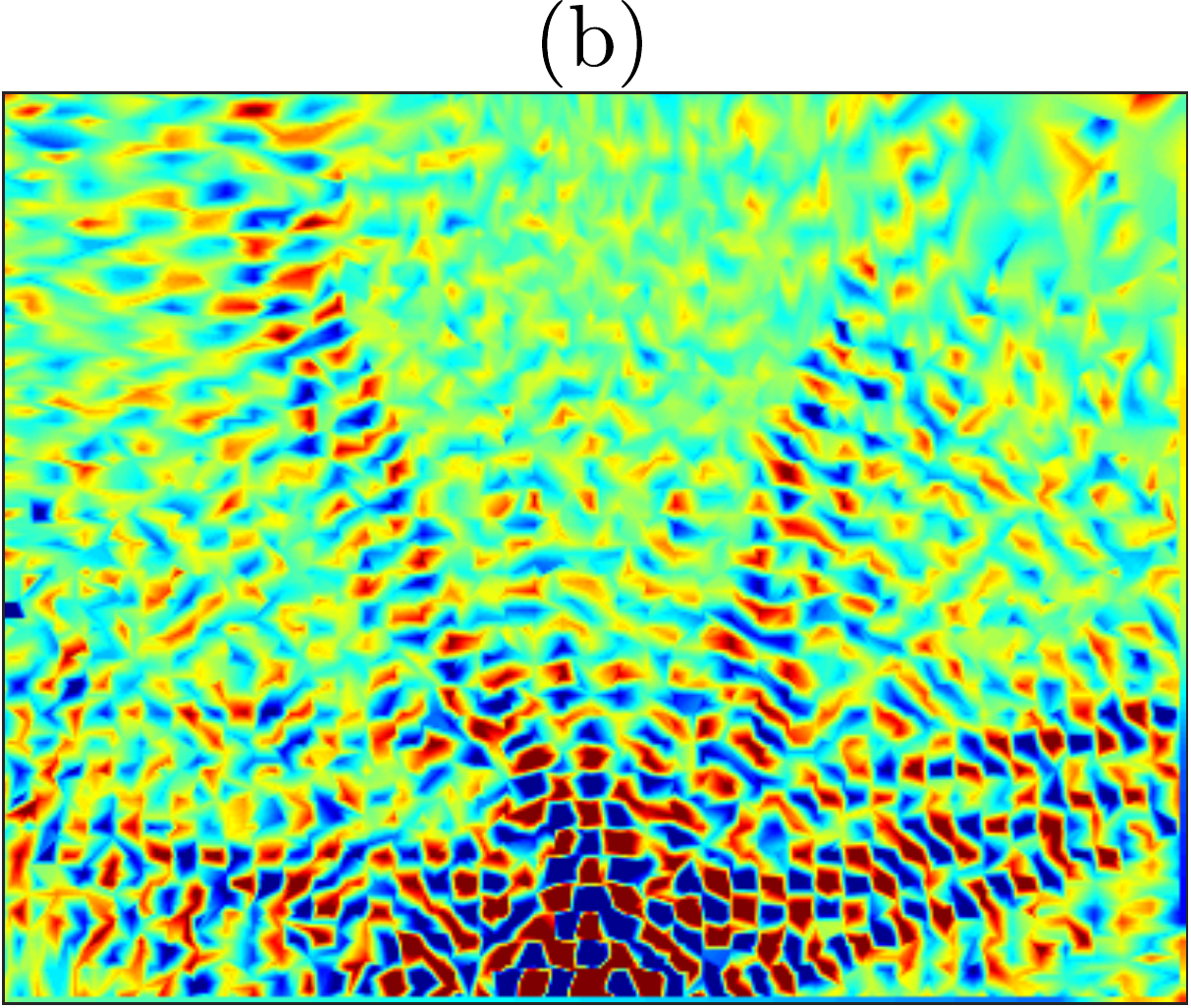} \qquad
\includegraphics[scale=0.375, trim = 0mm 9mm 0mm 0mm]{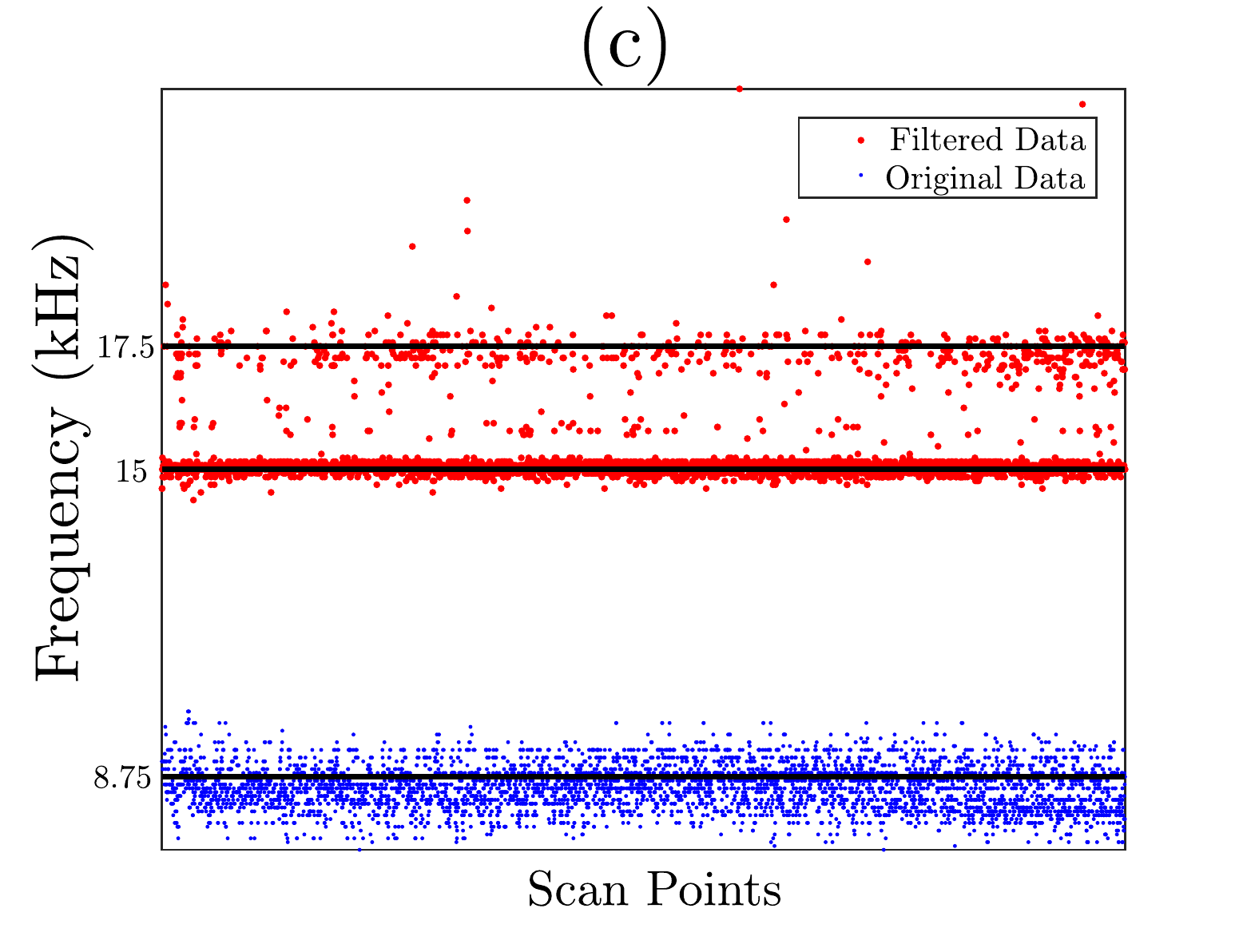} \\\vspace{0.25in}
\hspace{-0.175in}\includegraphics[scale=0.375, trim = 5mm 0mm 0mm 0mm]{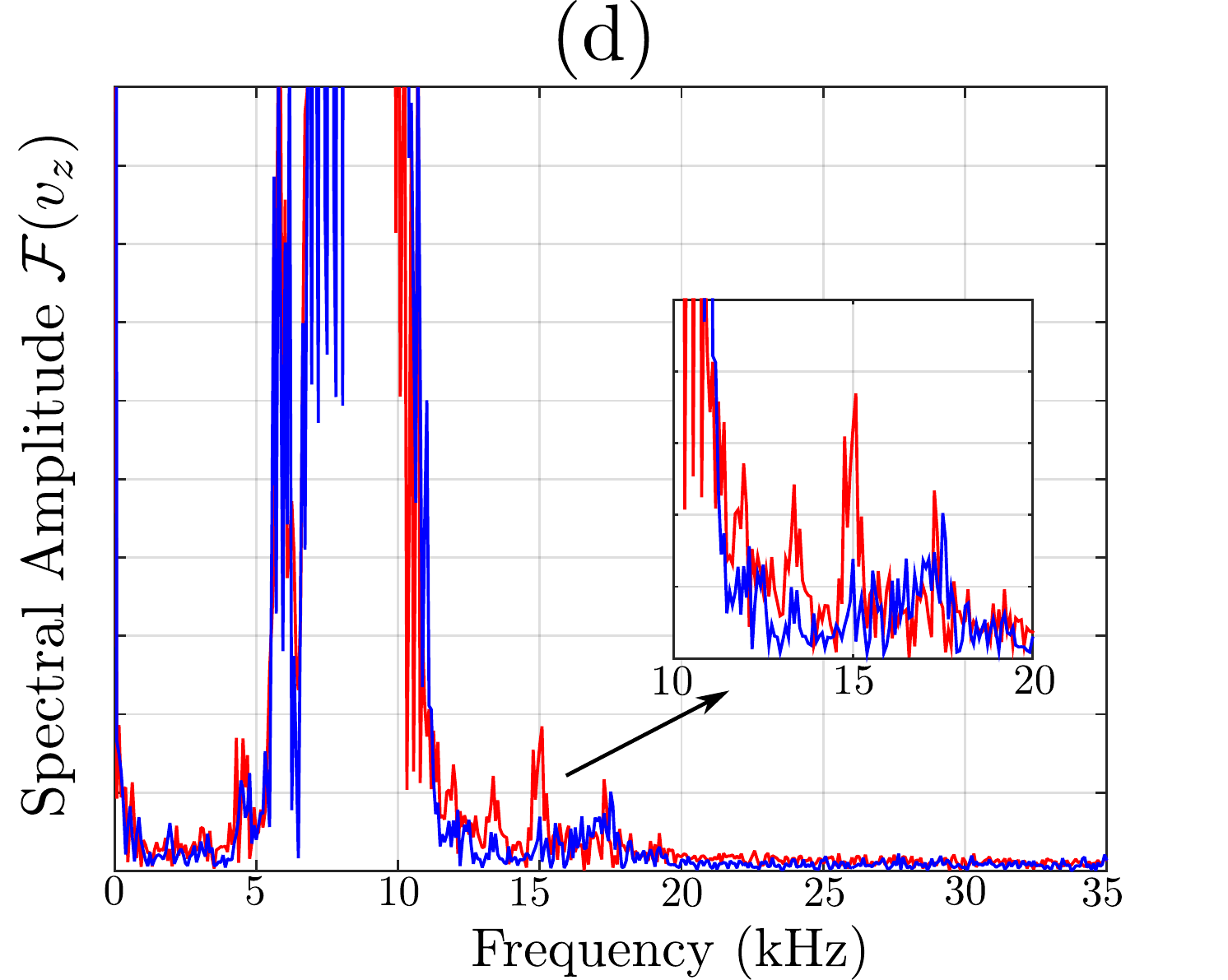} \qquad 
\includegraphics[scale=0.375, trim = 12mm 0mm 0mm 0mm]{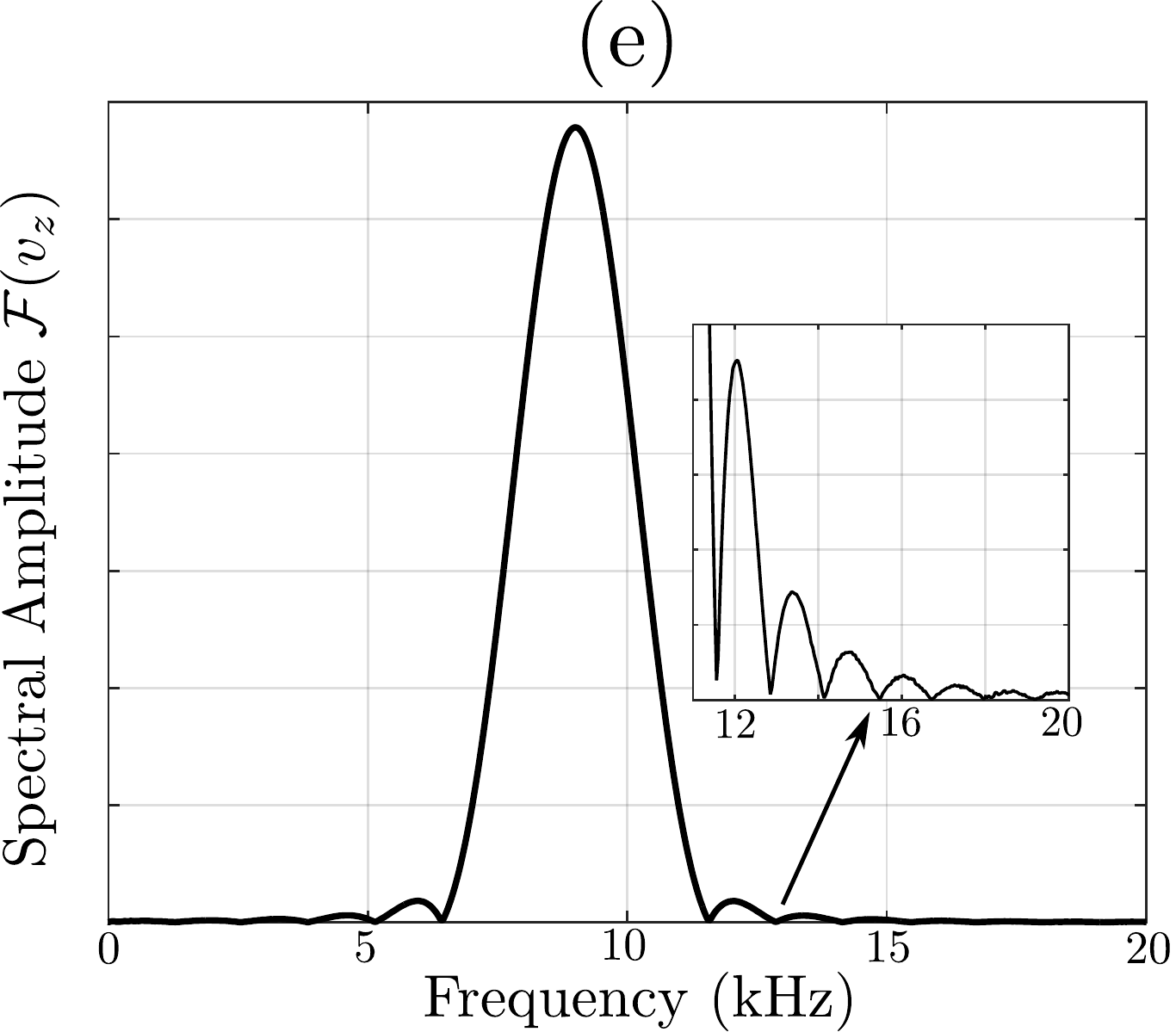} \qquad 
\includegraphics[scale=0.4, trim = 8mm 0mm 0mm 0mm]{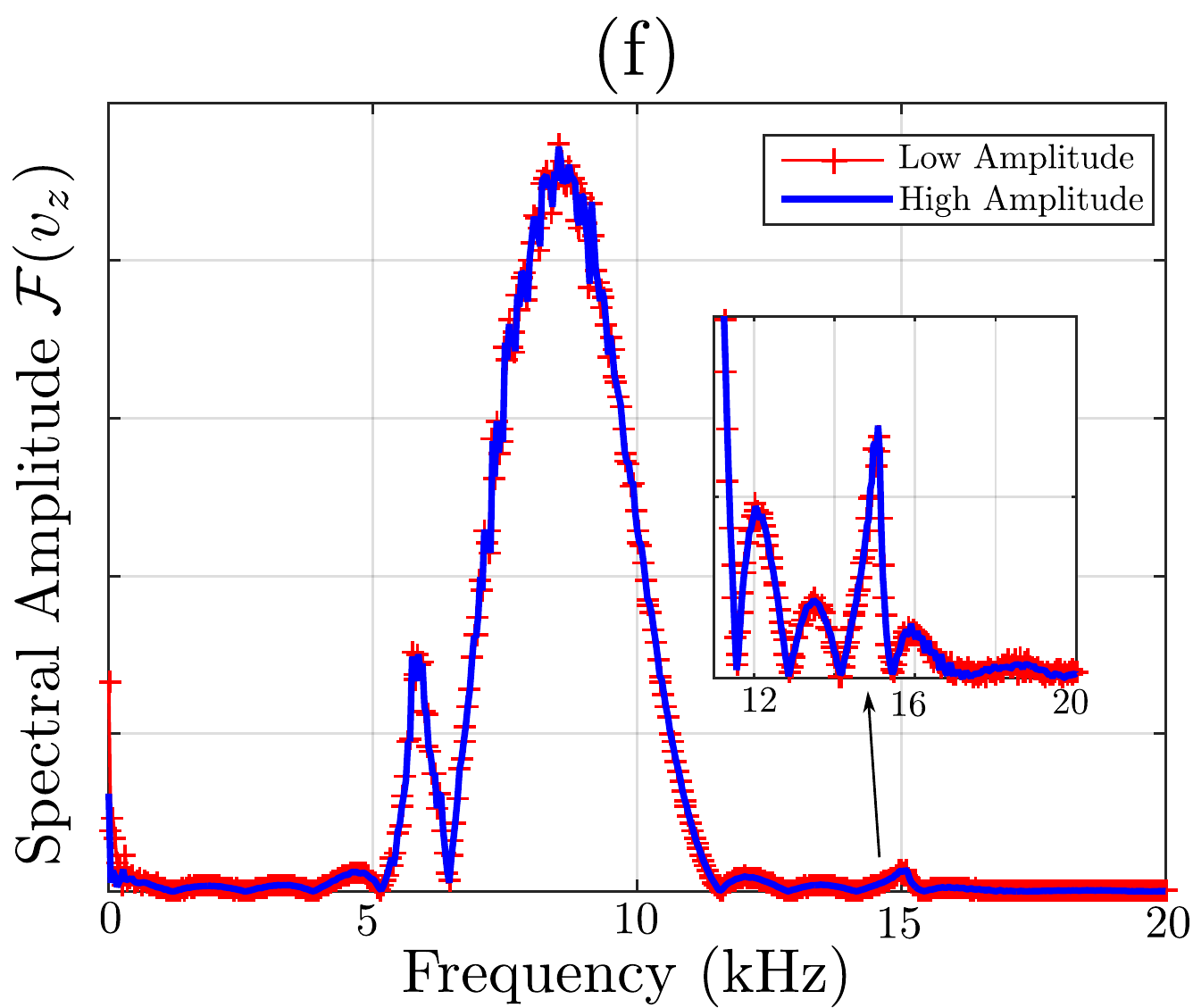}
\caption{(a) Snapshot of the experimental wavefield for an excitation at $\omega = 9$ kHz. (b) Snapshot of the wavefield obtained upon application of a band-pass filter to the full wavefield data. (c) Spectral signature of the full and filtered wavefield data for all scan points. The unfiltered data has one dominant spectral component at the excitation frequency, while the filtered data depicts two equally dominant harmonic components. (d) Frequency spectrum at two sample scan points, confirming the presence of two competing higher harmonic components (at $15$ kHz and $17.5$ kHz). (e) Spectral content of the input burst excitation at $9$ kHz. (f) Frequency spectrum measured at the point of excitation for two different input amplitudes. } 
\label{fig3}
\end{figure*}
In order to establish nonlinear effects in the Aluminum sheet, the amplifier (Bruel \& Kjaer Type $2718$) which powers the electrodynamic shaker is set to maximum gain, and the initial amplitude of excitation is carefully chosen such that the maximum current supplied to shaker does not exceed its operating limit. The out-of-plane velocity time histories measured at the scan points are then utilized to reconstruct the transient spatial response. A snapshot of this wavefield is shown in fig.~\ref{fig3}(a), where we observe two wave fronts along the diagonal directions emanating from the point of excitation, as predicted by the group velocity contours (for the lowest mode) in fig.~\ref{fig2}(b). \newline
Since the nonlinear features are expected to be much smaller in magnitude than the linear component of the response, we apply a band-pass filter (centered at twice the excitation frequency) to separate any potential higher harmonic components. A snapshot of this filtered wavefield is shown in fig.~\ref{fig3}(b), and reveals an orthotropic wave pattern with beams along the horizontal and vertical directions, conforming to the group velocity contours predicted for the second flexural mode at $18$ kHz (insert in fig.~\ref{fig2}b). It is now imperative to establish that the observed features are indeed the signature of nonlinearity (as opposed to other high-frequency non-idealities in the system), and that the observed nonlinear effects are due to the structural system and not due to other possible sources of nonlinearity (e.g. electronics, excitation system, contact between actuator and structure, etc.). \newline%To this end, we carefully monitor the time histories (and the related spectral components) of several critical points in the structure as well as in the actuation line., complementary to that observed in Fig. 3(a) and conforming to the group velocity contour predicted for the second flexural mode at 2omega = 18kHz (insert in Fig. 2b).. While the spatial profile does feature directional characteristics along the horizontal and vertical directions of the lattice, similar to the group velocity contours shown in fig.~\ref{fig2}(b) (at $18$kHz), these features need to be ratified as a manifestation of nonlinear deformation mechanisms in the structure. %its spectral signature needs to be estimated and needs to be observed as anin order to determine the presence of higher harmonic features.
 To this end, we evaluate the Discrete Fourier Transform (DFT) of the velocity time history for each scan point, and determine the frequency corresponding to the maximum spectral amplitude. We perform this analysis for the original as well as the band-pass filtered data, and the results are plotted in fig.~\ref{fig3}(c). As expected, the maximum spectral amplitude for the measured data is observed in the neighborhood of the prescribed excitation frequency ($9$ kHz). On the other hand, the maximum spectral amplitude for the filtered data is distributed around two frequencies at approximately $15$ kHz and $17.5$ kHz; this implies that the filtered component has two equally dominant harmonic components. This is further confirmed for two sample scan points in the frequency spectrum of fig.~\ref{fig3}d. While one of the frequencies does correspond to twice the excitation frequency, and can probably be attributed to nonlinear effects, the source of the other high-frequency component is not evident at this stage. As hinted above, a likely source of these effects could be the excitation system at large(function generator$\rightarrow$amplifier$\rightarrow$shaker/stinger$\rightarrow$specimen). \newline
In order to resolve this ambiguity, we monitor the spectral content of the time history measured directly at the excitation point (at the scan point corresponding to the exact location where the shaker is connected to the structure through the stinger). This measurement is repeated for different amplitudes of excitation, and the results of two extreme cases (low and high amplitude) are compared in fig.~\ref{fig3}(f). Note that the spectral signature of the input excitation supplied to the amplifier, shown in fig.~\ref{fig3}(e) for comparison, features a mainlobe centered around the prescribed frequency and a cascade of sidelobes of exponentially reducing amplitude - a characteristic commonly associated with burst excitations. In the measurements made at the scan point, which essentially capture the initial excitation supplied to the structure, some sidelobes feature stronger amplitude than the others. In particular, we observe a relative spike in the sidelobe centered at $15$ kHz, which provides a rationale for the frequency component observed at $15$ kHz in the filtered wavefield. % (note that there is also a spike in the sidelobe around $7$ kHz, which can be neglected for this study).
Nevertheless, the amplitude of these sidelobes does \textit{not} depend on the amplitude of the excitation supplied to structure, as shown in fig.~\ref{fig3}(f). This implies that the interaction between the shaker and the structure can be categorized as a linear process. In other words, the spectral component observed at $15$ kHz is attributed to a non-ideal, yet linear transduction process associated with the actuation system and/or the interaction of the stinger with the structure. \newline 
%\begin{figure*}[!htb]
%\includegraphics[scale=0.34]{BurstSpec} \qquad 
%\includegraphics[scale=0.375]{HighLowComp} \qquad
%\includegraphics[scale=0.375]{EdgeCenterComp} \\
%\caption{(a) Spectral signature of burst excitation, centered at $9$kHz. (b) Spectral signature of measured velocity wavefield in the out-of-plane direction at the point of excitation for two different amplitudes of excitation. (c) Comparison of the spectral signature of velocity wavefield in the out-of-plane direction measured at the point of excitation for two different excitation points. The relative amplitude of the sidelobes is shown in the inset in each figure.}
%\label{fig4}
%\end{figure*}
\begin{figure}[!htb]
\includegraphics[scale=0.375, trim = 5mm 0mm 0mm 0mm]{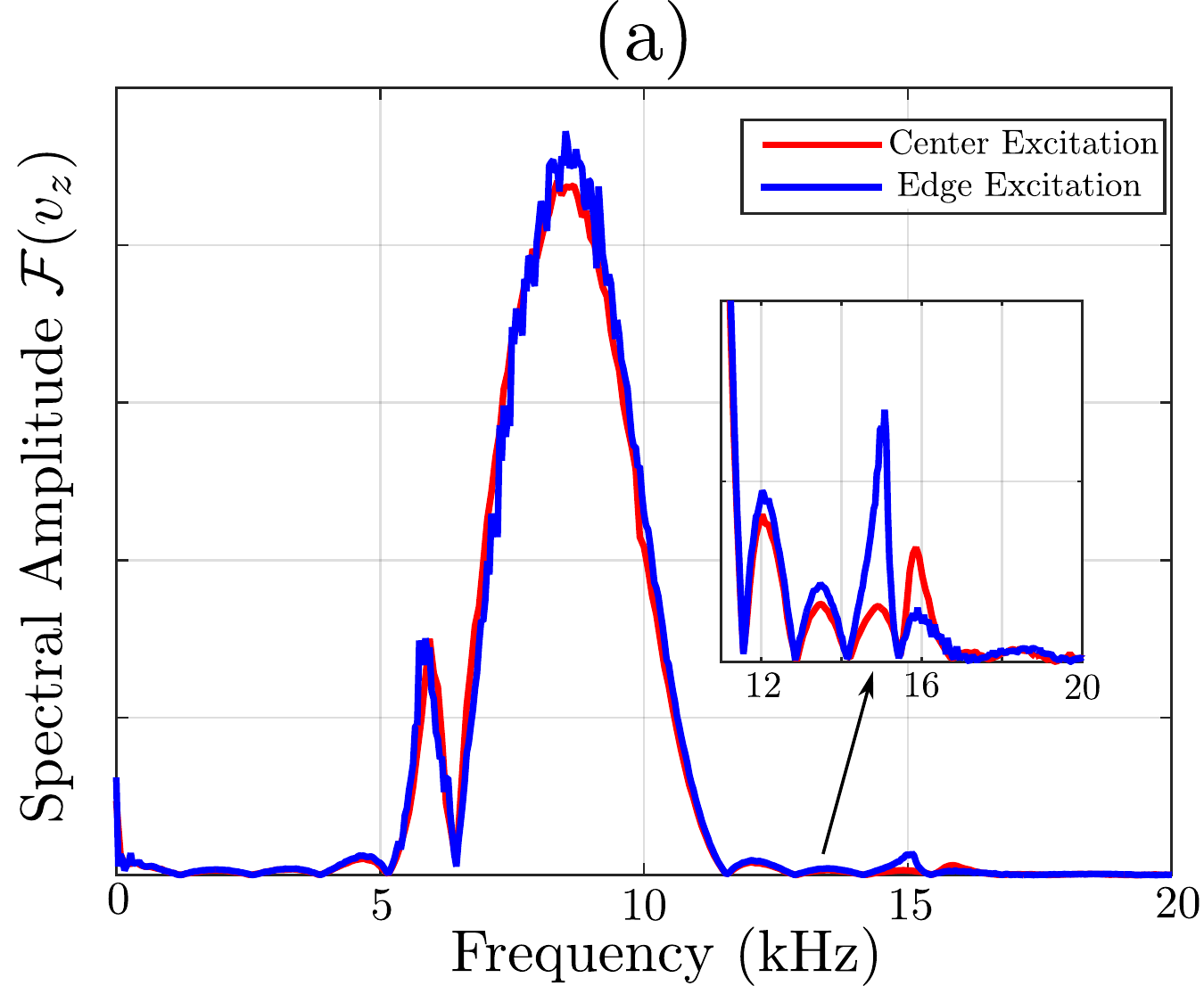} \\ \vspace{0.15 in} 
\includegraphics[scale=0.375, trim = 0mm 0mm 0mm 5mm]{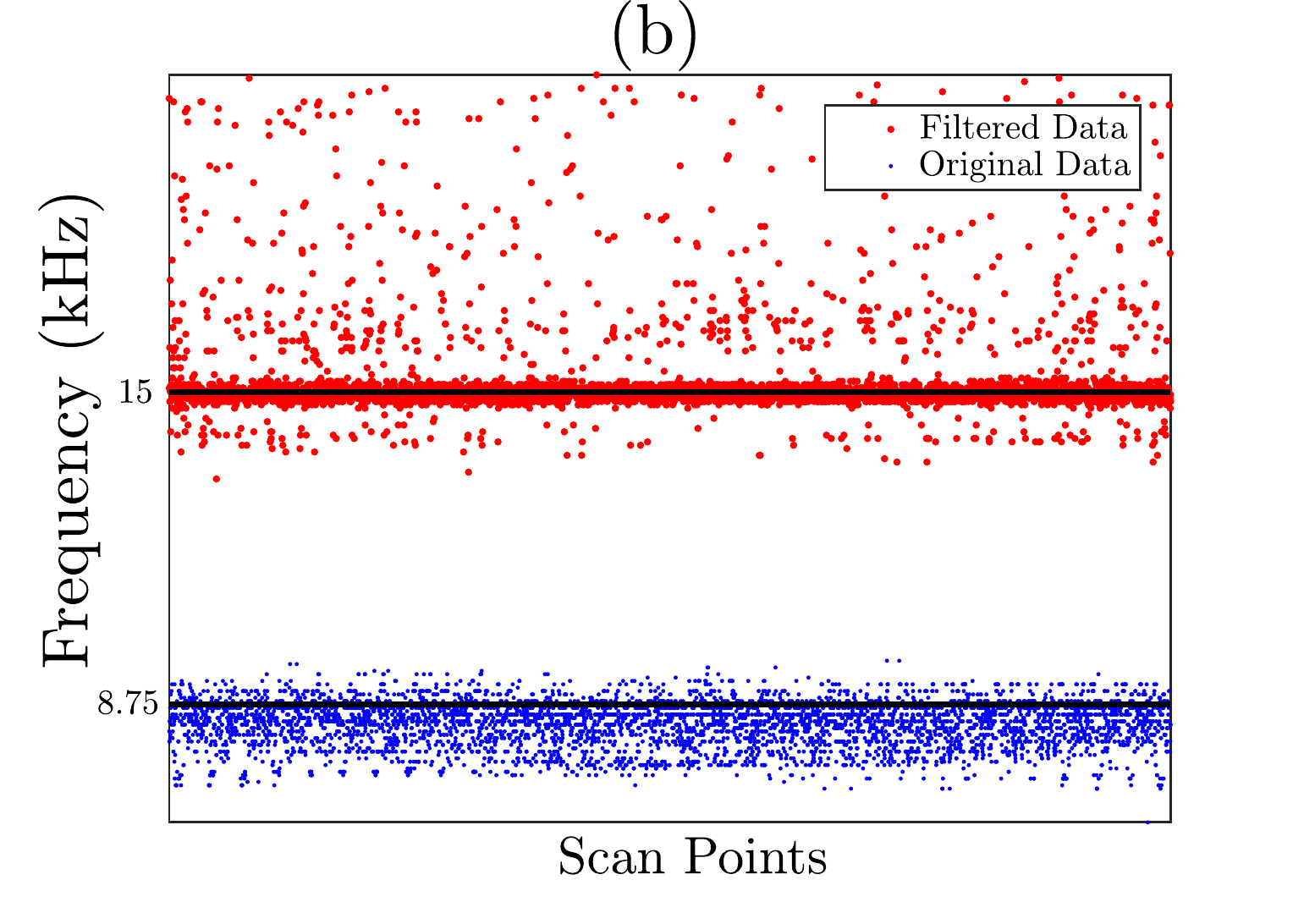} 
\caption{(a) Comparison of the spectral content of the time histories measured at the point of excitation for two different excitation scenarios (edge vs. center). (b) Spectral signature of the full and filtered data for all the scan points, for the $9$kHz excitation applied in the center of the specimen. In this case, the signature of the harmonic at twice the excitation frequency is not observed.}
\label{fig4}
\end{figure}
Moreover, in the measurements at the excitation point, no spike is observed in the sidelobe at twice the excitation frequency. In contrast, as the excitation propagates in the structure, the amplitude of the second harmonic component becomes comparable to the one at $15$ kHz, as indicated by the cluster of points in the neighborhood of $17.5$ kHz in fig.~\ref{fig3}(c). This suggests that the second harmonic component manifests solely due to nonlinear deformation mechanisms activated in the structure. If this conjecture is valid, then this component must display a dependence on the initial amplitude of excitation. We test the amplitude-dependency of the nonlinearly-generated higher harmonic by moving the source of excitation to the middle of the specimen, while keeping the input amplitude constant. By exciting the structure in the center, we effectively reduce the amplitude of the wave in each quadrant by a factor of two, as the energy is split into four quadrants, as opposed to just two (for the case of the edge-excitation). As the nonlinearly-generated harmonic depends on the square of the amplitude of the fundamental harmonic, its amplitude should reduce by a factor of four\cite{Ganesh2016}, while the amplitude associated with the sidelobes will only be halved. \newline% Therefore, we repeat the acquisition and analysis for the center-excitation, and employ the same band-pass filter to separate the higher harmonic components.
In fig.~\ref{fig4}(a), we first compare the spectral component of the velocities measured at the excitation point for the edge- and center-excitations, where we observe that the input excitation in both cases is almost identical (the center excitation has a spike in the amplitude of the sidelobe at $16$ kHz instead of $15$ kHz). This implies that changing the location of the excitation does not significantly affect the transduction process, and the fundamental excitation supplied to the structure essentially remains the same. In contrast, when we repeat the spectral analysis using the measured and filtered velocity time histories, we observe that the high-pass filtered component only consists of a single dominant frequency, corresponding to the intrinsic sidelobe of the input excitation (shown in fig.~\ref{fig4}b). This allows us to definitively establish that the second harmonic component observed in fig.~\ref{fig3}(c) does indeed display amplitude-dependent characteristics, and we can conclude that the propagating wavefield does contain a component due to nonlinear structural mechanisms. \newline
\begin{figure}[!htb]
\includegraphics[scale=0.43, trim = 5mm 0mm 0mm 0mm]{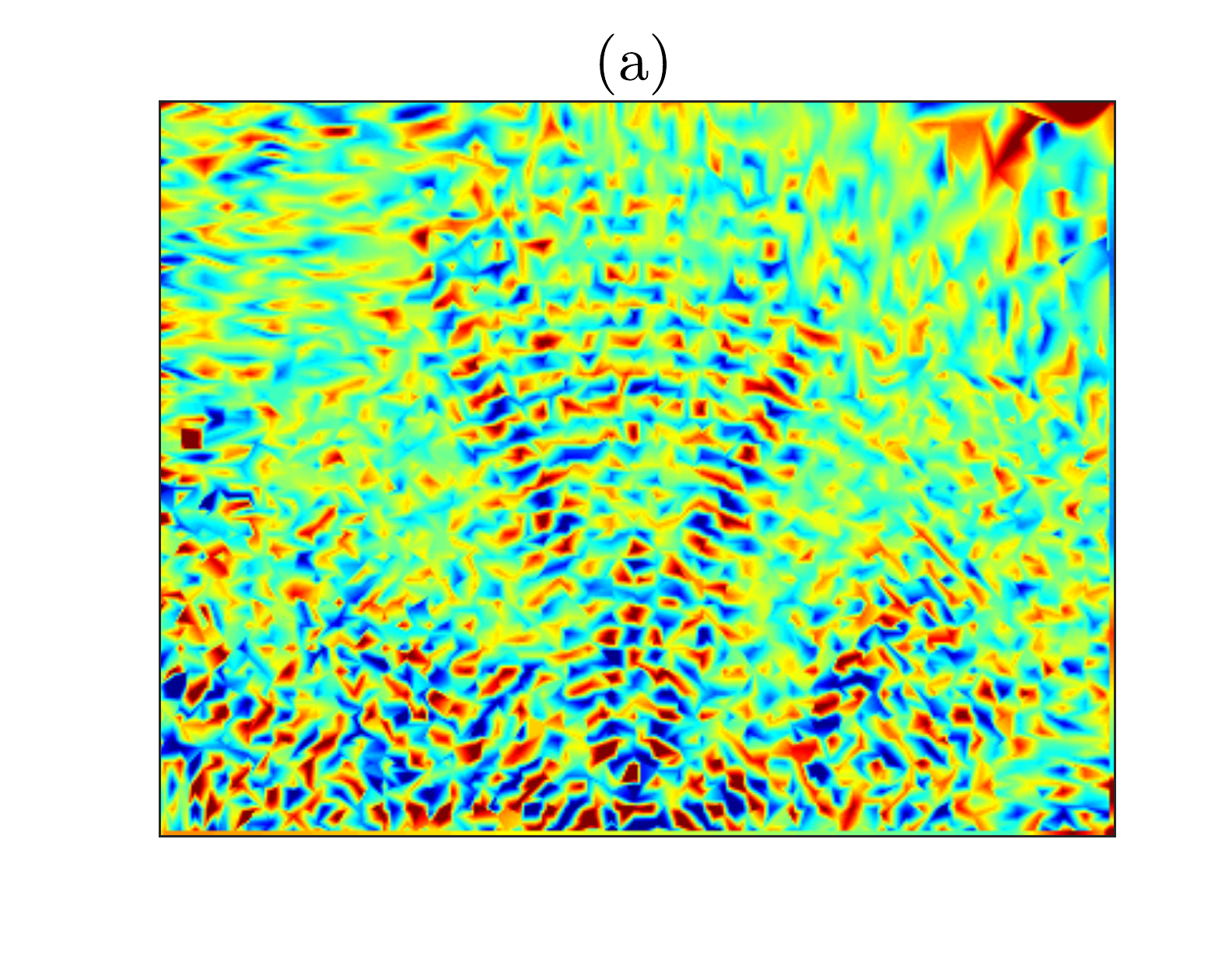} \\\vspace{-0.1in}
\includegraphics[scale=0.48]{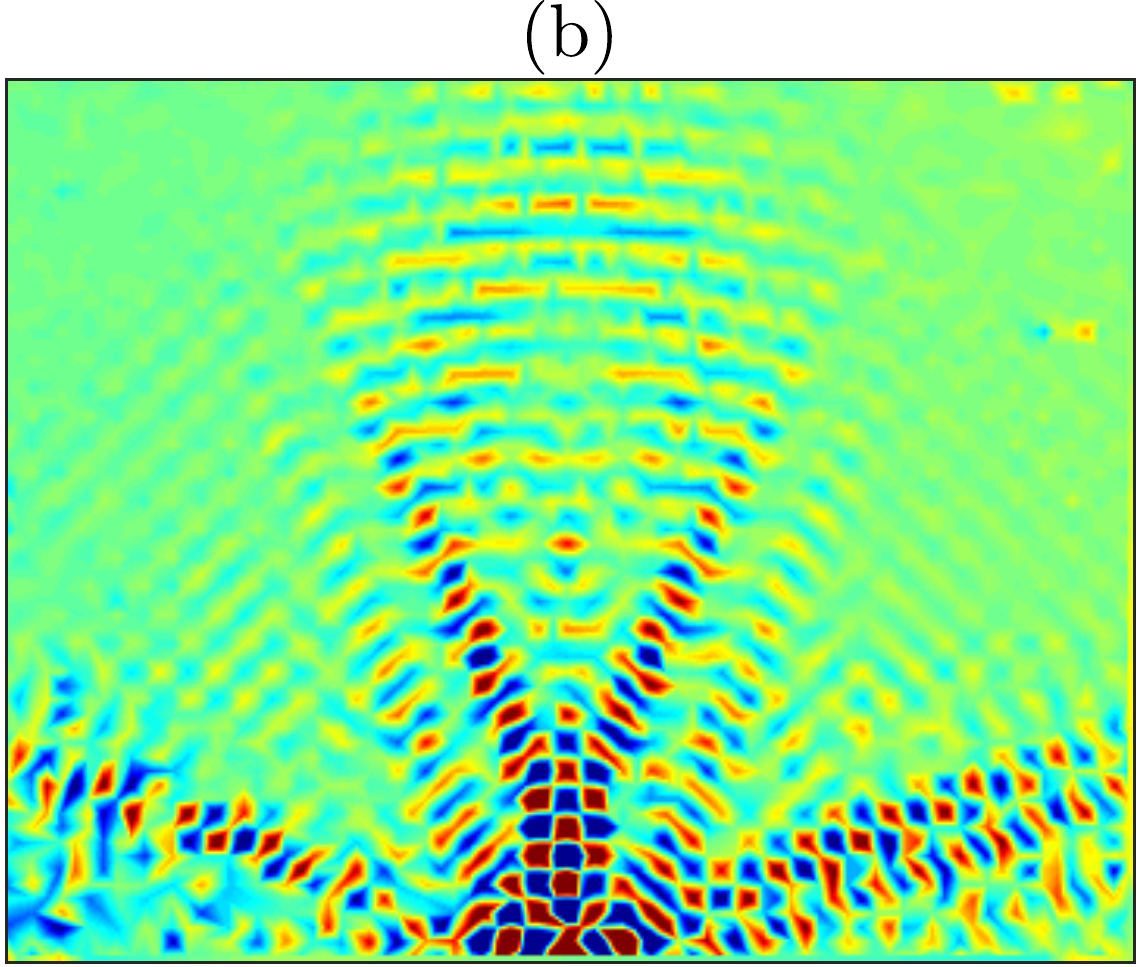} 
\caption{(a) Wavefield obtained by isolating the spectral component centered around $17.5$kHz for a large-amplitude excitation centered at $9$kHz. (b) Snapshot of the wavefield obtained for a small-amplitude excitation centered at $\omega = 17$kHz.}% The patterns of the two wavefields are remarkably similar.}
\label{fig5}
\end{figure}
Finally, we reconsider the case of the edge-excitation (fig.~\ref{fig3}(a-b)) to further investigate the spatial characteristics of the nonlinearly-generated second harmonic and to distill it more effectively from the competing non-idealities that co-exist in the same frequency interval. To this end, we employ a finer band-pass filter, designed to exclude the sidelobes at $15$ kHz; a snapshot of this new filtered wavefield is shown in fig.~\ref{fig5}(a). In order to confirm the meaningfulness of the observed spatial patterns, the experimental acquisition is repeated for a low-amplitude edge-excitation at $17$ kHz (since the frequency of the electrodynamic shaker is limited at $20$kHz, the tone-burst is centered at $17$kHz to ensure that the mainlobe is completely excited in the structure), and a snapshot of the acquired fundamental response is shown in fig.~\ref{fig5}(b). %, which captures the caustics observed in fig.~\ref{fig2}(b) (in the vertical direction) accurately. 
The morphologies of the two wavefields are remarkably similar and both feature the presence of the vertical caustics predicted by unit cell analysis (the noisy features can be attributed to other contributions associated with nonlinear wave propagation\cite{Ganesh2016}, as well as to additional experimental noise). While fig.~\ref{fig3}(b) also shows a wavefront propagating along the horizontal and vertical directions, the angle spanned by the wavefront in the vertical direction is larger than that observed in fig.~\ref{fig5}(a-b), which can be attributed to the presence of the other frequency component at $15$ kHz. Fig.~\ref{fig5}(a) ultimately confirms that the harmonic component generated due to nonlinear mechanisms features characteristics that are described by those of the corresponding linear system. \newline%, and is not associated with any nonlinear processes. \newline
In conclusion, we have experimentally investigated nonlinear wave propagation in lattice structures and demonstrated that the interplay of dispersion, nonlinearity, and modal complexity involved in the generation and propagation of higher harmonics gives rise to secondary wave packets with characteristics that conform to the dispersion relation of the corresponding linear structure. As a result, by simply analyzing the linear wave characteristics of a periodic structure, we can design a wide spectrum of modal functionalities that can be activated through nonlinear mechanisms. While the nonlinear features that have been measured in this example are indeed much smaller than the linear component, and can be probably neglected for all practical applications, these results provide proof of concept for a methodology that can be easily extended to systems with different materials (for example, soft polymeric materials such as ABS\cite{CelliLego}), where the effects of nonlinearity and the associated functionalities could be markedly more pronounced. \newline
The authors acknowledge the support of the National Science Foundation (CAREER Award CMMI-1452488).

\bibliography{paper}%
\end{document}